# A DENSENET-BASED METHOD FOR DECODING AUDITORY SPATIAL ATTENTION WITH EEG

*Xiran Xu [1,3], Bo Wang[1,3], Yujie Yan[2,3], Xihong Wu[1, 3], Jing Chen[1,2,3]*

[1] Speech and Hearing Research Center, School of Intelligence Science and Technology, Peking University
[2] National Biomedical Imaging Center, College of Future Technology, Peking University
[3] National Key Laboratory of General Artificial Intelligence, China
janechenjing@pku.edu.cn

## ABSTRACT

Auditory spatial attention detection (ASAD) aims to decode the attended spatial location with EEG in a multiple-speaker setting. ASAD methods are inspired by the brain lateralization of cortical neural responses during the processing of auditory spatial attention, and show promising performance for the task of auditory attention decoding (AAD) with neural recordings. In the previous ASAD methods, the spatial distribution of EEG electrodes is not fully exploited, which may limit the performance of these methods. In the present work, by transforming the original EEG channels into a two-dimensional (2D) spatial topological map, the EEG data is transformed into a three-dimensional (3D) arrangement containing spatial-temporal information. And then a 3D deep convolutional neural network (DenseNet-3D) is used to extract temporal and spatial features of the neural representation for the attended locations. The results show that the proposed method achieves higher decoding accuracy than the state-of-the-art (SOTA) method (94.3% compared to XANet's 90.6%) with 1-second decision window for the widely used KULeuven (KUL) dataset, and the code to implement our work is available on Github:
   https://github.com/xuxiran/ASAD_DenseNet

## 1. INTRODUCTION

In a complex auditory scenario where multiple speakers are talking at the same time, i.e. cocktail party [1], humans are able to selectively attend to and maintain the focus on one speaker while ignoring other speakers and noise. Neuroscience studies indicate that it is possible to identify the attended speech with listeners' EEG since the cortical speech-envelope tracking of the attended stream was significantly enhanced by auditory selective attention [2-4]. The corresponding technology developed to decode the attended object with EEG/MEG is named as auditory attention decoding (AAD), which is potentially applicable for the realization of cognitively-controlled, or neuro-steered, hearing equipment [5-6].

In the general AAD algorithms, known as the stimulus reconstruction approaches [2,4], a speech envelope was constructed from EEG by a decoder, and the correlations between the constructed speech envelope and the attended/unattended speech envelope were used to identify the attended object, with which the higher correlation indicated the attended one [2,4]. On the other side, the auditory spatial attention detection (ASAD) methods [7-13] were also developed to decode the attended object, since the locus of auditory spatial attention was neurally encoded as brain lateralization when the target and the competing auditory streams were spatially separated [14-16]. Compared to the stimuli-construction methods, the ASAD methods have two main advantages: 1) it can identify the attended sound stream directly from the user's EEG, i.e., without requiring access to demixed and clean speech signals; (2) it can operate accurately on short time-scales (within 1–5 s) [7]. In this work, we focused on improving the performance of ASAD.

The existing ASAD methods [7-13] can be divided into two types: the traditional decoders consisting of feature extraction frontend and pattern classification backend [7,9], and the DNN-based methods [11,13]. The traditional decoders were usually based on second-order statistics of the EEG data [7,9]. For example, the common spatial pattern (CSP) [7] was a data-driven filtering approach based on the EEG covariance structure, aiming to extract the most discriminative spatial patterns from the EEG signals. The Riemannian geometry-based classifier (RGC) [9] was further developed to exploit the symmetric positive definite property of the covariance matrices. Both CSP and RGC methods could outperform the stimulus reconstruction approach [2,4] with KULeuven (KUL) dataset [17], i.e. RGC achieves about 80% and 85% with the window length of 2 and 5 seconds, respectively. However, the performance of these methods was still limited with short decision windows, i.e., 1-second windows.

Since ASAD is a nonlinear classification problem, deep learning is expected to provide a better solution [11-13,18,19]. Vandecappelle et al. [12] proposed a CNN-based ASAD model and the decoding accuracy was around 82% for 1-second decision windows in the KUL dataset. The attention-based neural networks were introduced in recent works [11,13,18], and the decoding accuracy was achieved at 90.6% (SOTA, XAnet, [18]) with 1-second decision windows in the KUL dataset.

However, the spatial distribution of EEG electrodes was not fully exploited in the above methods. It has been reported that introducing the spatial distribution of EEG electrodes, by transforming the general input of sequential EEG channels into a 2D spatial topological organization, can extract the spatial and temporal features simultaneously [20-24]. In ASAD, Zhang et al. [13] proposed a learnable spatial mapping method combined with the spatial attention mechanism to improve the efficacy of utilizing the spatial distribution of EEG electrodes. Jiang et al. [25] also projected the EEG electrode channels onto a 2D plane and used CNN and convolutional-long-short-term-memory (ConvLSTM) to extract features. Although these two works introduced the spatial distribution of EEG electrodes, their performance was still worse than the model proposed by Pahuja et al. (XANet [18]), probably due to the powerful capability to extract spatial-temporal features for XANet.

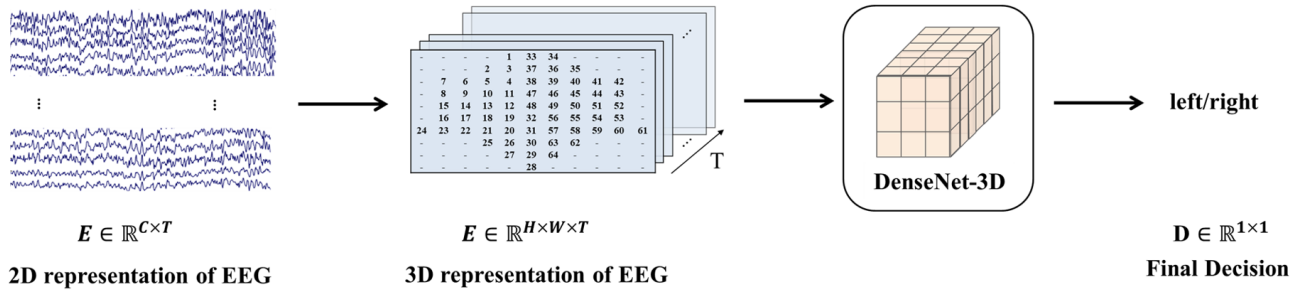

**Fig. 1.** The architecture of ASAD-DenseNet

Compared to traditional models, such as CNN and ConvLSTM, models utilizing residual connections [26] and dense connections [27] have been validated to aid network training and prevent feature loss as the network depth increases. We assumed that when introducing the spatial distribution of EEG electrodes, the utilization of residual connections [27] and dense connections [26] could be able to extract more robust spatial-temporal features [28].

DenseNet-3D, a 3D deep convolutional neural network (3D ConvNets), has shown effective in extracting spatial-temporal features across various fields such as video understanding [29], EEG emotion recognition [28], and electrocorticogram (ECoG) movement classification[30]. In order to further extract features and improve the decoding accuracy of ASAD, ASAD-DenseNet was proposed to decode auditory spatial attention in this work.

## 2. ASAD-DENSENET

### 2.1. 3D Representation of EEG

The architecture of ASAD-DenseNet is illustrated in Figure 1. Let $E \in \mathbb{R}^{C \times T}$ be the EEG signals for a decision window, with C channels and T time samples. The ASAD-DenseNet takes E as the input and makes auditory spatial attention decisions. To get the 3D representation of EEG, the channel indexes are converted from 1D (with the number of C) to 2D (H×W) in the same manner used in [20]. The EEG signals are transformed from $E \in \mathbb{R}^{C \times T}$ into $E \in \mathbb{R}^{H \times W \times T}$ using the topological location of EEG channels in space and the blank positions are filled with zeros [20-24], where H and W represent the spatial (channel) dimension, shown as the input of Figure 2, and T represents the temporal dimension.

### 2.2. DenseNet-2D

Considering that 3D ConvNets have many parameters that makes it hard to train [31], a DenseNet-2D model was proposed to initialize the parameter of DenseNet-3D using bootstrapping [31]. At specific sampling time of 3D representation of EEG $E \in \mathbb{R}^{H \times W \times T}$, corresponds to $E_t \in \mathbb{R}^{H \times W}$, which is a 2D topography and as the input of the DenseNet-2D. The architecture of DenseNet-2D is shown in Figure 2, which is the simplified version of DenseNet [27]. The first Convolution layer is BN-ReLU-Conv(1×1)-BN-ReLU-Conv(3×3), containing the Batch Normalization (BN), rectified linear units (ReLU) and Convolution (Conv). Max Pool with stride 2 is then used and each side of the input is zero-padded by one unit. The key part is the four dense blocks and the three transition layers. Each dense block contains four BN-ReLU-Conv(1×1)-BN-ReLU-Conv(3×3) layers. The input and output of a specific layer in the dense block are concatenated as the input of the next layer. The transition layers between two contiguous dense blocks contain BN-ReLU-Conv(1×1) and 2×2 average pooling with stride 1 without any padding. At the end of the last dense block, a global average pooling is performed, and then a fully connected layer followed by a softmax classifier is attached. The output of the classifier is a binary decision, and the two values, 0 and 1, indicate the attended location with the EEG input at time $t$, $E_t$, decoded as "left" or "right", respectively.

### 2.3. DenseNet-3D

To extract temporal and spatial features of EEG signals, the DenseNet-2D was transferred to DenseNet-3D by inflating its filters and pooling kernels [31]. In this work, each convolution layer in DenseNet-2D was inflated to BN-ReLU-Conv(1×1×1)-BN-ReLU-Conv(3×3×1), and the transition layer containing 1×1 convolution followed by 2×2 average pooling was inflated to the one containing 1×1×1 convolution followed by 2×2×7 average pooling with stride (1,1,3).

The parameters of DenseNet-3D were also initialized from DenseNet-2D using bootstrapping, which can be achieved by repeating the weights of the 2D filters N times along the temporal dimension and rescaling them by dividing by N because of linearity. More details about bootstrapping can be found in [31].

## 3. EXPERIMENTS

### 3.1. Dataset

The ASAD experiments were conducted on the KUL dataset [17], which was used for ASAD in many previous studies [7-13]. This dataset contains the EEG recordings from 16 normal hearing subjects and 48 minutes for each of them. During the EEG recording, the subject was required to pay attention to one of the two simultaneously active competing speakers, and the two speakers were located at ±90° along the azimuth direction, corresponding to the left and right location, respectively. The EEG was recorded using a 64-channel BioSemi Active Two system. More details are available in [17].

### 3.2 Data Preprocessing

The EEG data was down-sampled to 128 Hz, which was as same as previous work [11,12,18]. According to the analysis of the filter band importance in the previous studies [12], the β-band was the most useful EEG frequency band to decode the directional focus of attention [9]. As a result, the EEG data was bandpass filtered between 14 and 31 Hz by an 8th order Butterworth filter. Finally, the EEG data were normalized to ensure zero mean and unit variance across channels and along time. Because the proposed ASAD-

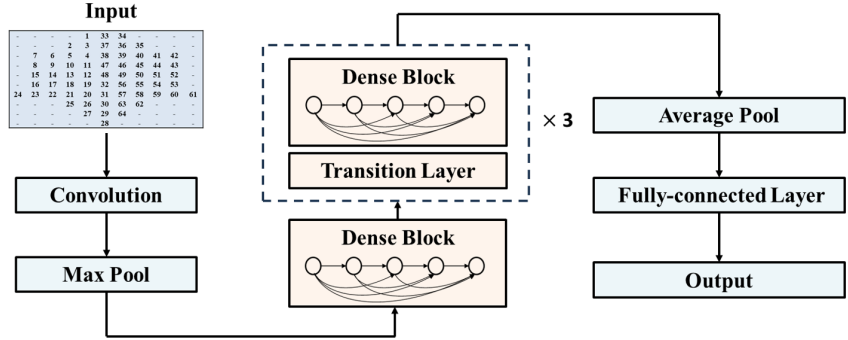

**Fig. 2.** The architecture of DenseNet-2D. The architecture of DenseNet-3D can also be described by this figure by modifying the input from specific timepoint $E_t \in \mathbb{R}^{H \times W}$ to EEG data $E \in \mathbb{R}^{H \times W \times T}$.

DenseNet was a data-driven solution, that was expected to function in an end-to-end manner, no artifacts removal operation was involved in the data processing. This could facilitate the implementation of real-time BCI systems, such as neuro-steered hearing aids.

### 3.3 Network Configuration

Four decision windows were analyzed in this work, i.e., 1, 2, 5, and 10 seconds. For each of four window length conditions, the performance of the proposed methods was evaluated using 5-fold cross-validation (CV) over all windows [11]. Each decision window would output a decision, and the ASAD accuracy was calculated as the percent correct of the correct decisions among windows. The EEG data with the label of attended location were divided into five groups equally. Four of them were used as the training data, and the remaining group was used for the test. This process was repeated five times until all data were tested once. The training data was partitioned into training and validation sets with the ratio of 4:1. The training data were first used to train and validate DenseNet-2D, and then the parameters of DenseNet-2D were used to initialize the parameters of DenseNet-3D by bootstrapping [31]. Next, training data were used to train and validate the DenseNet-3D. Finally, the trained model was used to test the decoding accuracy.

The results in both subject-independent and subject-dependent conditions would be reported. In the subject-independent condition, all subjects' data were trained and tested using the same model, while in the subject-dependent condition, a separate model was trained and tested for each subject.

The neural networks were implemented with the Pytorch framework and trained on an NVIDIA RTX 3090 GPU. The Adaptive Moment Estimation (Adam) optimizer [32] was employed to minimize the cross-entropy loss function with the learning rate of $10^{-3}$.

### 3.3 Models

Experiments were conducted on four models to do ablation analysis with the subject-dependent decoder in subject-dependent condition. The first model (CNN-baseline) was a CNN model as the same as Vandecappelle et al. (2021), which used the 2D representation of EEG data $E \in \mathbb{R}^{C \times T}$ as input and used five independent 64 × 17 filters to extract feature. ReLU was used as the activation function after the convolution step followed by average pooling over the temporal dimension and the two fully connected layers.

A simple CNN-3D model was proposed to validate the effectiveness of providing spatial distribution of EEG electrodes. The input of this model was the 3D representation of EEG data $E \in \mathbb{R}^{H \times W \times T}$, where H = 10, W = 11 and T = 128 when the decision window was 1 second. In the CNN-3D model, twenty independent 5 × 5 spatial filters were used to extract features. ReLU was used as the activation function after the convolution step followed by average pooling over the temporal dimension and the two fully connected layers. The CNN-3D model was very similar to the CNN-baseline except it provided EEG data with spatial distribution of EEG electrodes and used simple spatial filters to extract features.

To validate the effectiveness of using deeper CNN, the third model was DenseNet-3D without bootstrapping. The last model was DenseNet-3D with bootstrapping, which was the proposed model in this work, which was used to validate the effectiveness of the operation of bootstrapping.

## 4. RESULTS AND DISCUSSION

### 4.1. Decoding Accuracy

As shown in Figure 3 (a), in the subject-dependent condition for the KUL dataset, the DenseNet-3D achieved an average decoding accuracy of 94.3% (SD: 5.7%), 95.9% (SD: 4.3%), 96.6% (SD: 3.9%), and 96.8% (SD: 3.6%) for the decision window length of 1,2,5,10 seconds, respectively. In general, a larger decision window provided a better result, consistent with the previous studies [9,11,13]. Compared with existing methods, ASAD-DenseNet outperformed previous models across all decision windows.

In the subject-independent condition, the DenseNet-3D achieved an average decoding accuracy of 94.3%, 95.6%, 96.1%, and 95.7% for the decision window length of 1,2,5,10 seconds, respectively. The results also revealed that, even in the subject-independent condition, DenseNet-3D achieved results similar to those in the subject-dependent condition, thus further demonstrating the model's ability to extract more robust and generalized spatial-temporal features.

### 4.2. Ablation Analysis

Figure 3(b) shows the ASAD accuracy of the four models across all subjects on the KUL dataset. For the 1-second decision window, the CNN-baseline model attained ASAD accuracy of 84.8% (SD: 10.4%). CNN-3D model outperformed the CNN-baseline model with an average improvement of 3.8% (88.6%, SD: 10.0%). A two-sample t-test revealed that there was a significant difference between the CNN-3D model and the CNN-baseline model (t = 3.6, p = 0.003) though they both used a simple CNN. This result validated the effectiveness of providing spatial distribution of EEG electrodes.

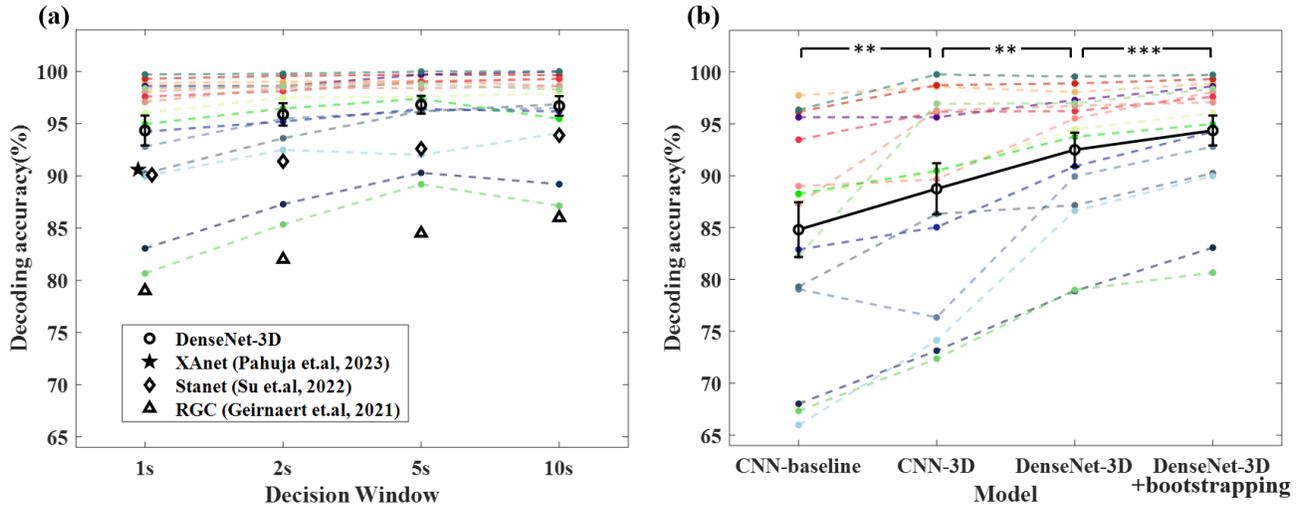

**Fig. 3.** Accuracy of different models among all subjects in the subject-independent condition. The dashed colorful lines represent the result of individual subjects, and the bold black line represents the average accuracy among subjects. (a) Accuracy of proposed models and some existing methods with 1,2,5 and 10-second decision window. XAnet only provides the decoding accuracy in 1-second decision window [18]. (b) Accuracy for each of the four models among all subjects with 1-second decision window. Error bars indicate the standard error. **$p < 0.01$, ***$p < 0.001$.

The DenseNet-3D without bootstrapping also outperformed the CNN-3D model, with an average accuracy of 91.8% (SD: 7.1%). The results revealed that there was a significant difference between the DenseNet-3D without bootstrapping and the CNN-3D model (t = 3.11, p = 0.008). The result validated that deeper CNN (DenseNet-3D) could further extract more robust spatial-temporal features and improve decoding accuracy.

ASAD accuracy of DenseNet-3D model with bootstrapping was 94.3% (SD: 5.7%). The results revealed that there was a significant difference between the DenseNet-3D model with bootstrapping and the DenseNet-3D model without bootstrapping (t =5.3, p < 0.001). The result suggested that DenseNet-3D with bootstrapping whose parameters were initialized from DenseNet-2D could get the highest decoding accuracy (94.3%, higher than 90.6% of SOTA XAnet [18]). This result verified the effectiveness of bootstrapping.

In summary, these results suggested the improvement of decoding accuracy could be attributed to three factors: providing spatial distribution of EEG electrodes, proposing a deeper CNN (DenseNet-3D) model, and the bootstrapping operation.

### 4.3. Limitations

The current model was only tested on the KUL dataset, which was widely used for AAD evaluation. Please notice that this dataset only included a simple scenario in which two speakers were located at ±90° along the azimuth direction, and it has been reported that the neural representation was the most discriminative for the spatial combination of ±90° with the multi-speaker setting, comparing to other spatial combinations [33]. Hence, the generalization to the real scenarios including diverse spatial combinations needs to be carefully treated. In addition, the efficiency of ASAD is dependent on the spatial separation between the attended target and the others, indicating the limitation for applying ASAD when the attended one is co-located with others. One possible solution is to combine ASAD with reconstruction-based methods to develop AAD methods for the more realistic scenarios and to improve decoding accuracy. Lastly, considering that DenseNet is a deep convolutional neural network, the computation cost should be much more than the baseline model. It is possible to reduce model parameters using model distillation.

### 5. CONCLUSION

In this work, a novel ASAD-DenseNet method is proposed. Compared to previous works, the EEG data is transformed into a three-dimensional (3D) arrangement that retains spatial-temporal information. The DenseNet-3D is employed to extract both temporal and spatial features of the neural representation for the attended locations. Experiments confirm the effectiveness of the proposed models and the decoding accuracy for 1-second EEG signal is 94.3% which is higher than that of SOTA. Both subject-dependent and subject-independent decoders show the promising advantages. It is worthy to explore the application of ASAD-DenseNet in more real auditory scenarios, such as the more complicated spatial combinations between the target and multiple interfering sounds, and with noise or reverberations.

### ACKNOWLEDGEMENTS

This work is supported by the National Key Research and Development Program of China (No.2021ZD0201503), a National Natural Science Foundation of China (No.12074012), and the High-performance Computing Platform of Peking University.